\documentclass[floats,floatfix,amssymb,prd,twocolumn,superscriptaddress,nofootinbib,preprintnumbers]{revtex4-2}

\usepackage{subcaption}
\usepackage{ragged2e}
\DeclareCaptionJustification{justified}{\justifying}
\makeatletter
\newcommand{\subsetsim}{\mathrel{\mathpalette\subset@sim\relax}}
\newcommand{\subset@sim}[2]{%
  \vtop{\offinterlineskip\m@th
    \ialign{\hfil##\cr
      $#1\subset$\cr\noalign{\kern0.5pt}\scalebox{0.9}{$#1\sim$}\cr
    }%
  }%
}
\makeatother

\usepackage{amssymb,amsmath,verbatim,mathtools,needspace,enumitem,etoolbox,graphicx,microtype,afterpage,bm}
\usepackage[normalem]{ulem}

\usepackage[dvipsnames, usenames]{xcolor}

\definecolor{linkcolor}{rgb}{0.0,0.3,0.5}
\usepackage{booktabs}
\usepackage[unicode, colorlinks=true, linkcolor=linkcolor, citecolor=linkcolor, filecolor=linkcolor,urlcolor=linkcolor, pdfusetitle]{hyperref}
\usepackage[all]{hypcap}
\usepackage[T1]{fontenc}
\usepackage[utf8]{inputenc}
\usepackage{tabularx}
\usepackage{float}
\usepackage{adjustbox}
\renewcommand{\arraystretch}{1.4}

\usepackage{multirow}
\usepackage{pifont}
\usepackage{lmodern}

\allowdisplaybreaks
\usepackage{tikz}
\usetikzlibrary{shapes.misc, positioning, arrows.meta}
\usepackage{framed}
\usepackage{hyperref}
\hypersetup{colorlinks, citecolor=bluscuro, linkcolor=black, urlcolor=bluscuro}
\definecolor{rossos}{cmyk}{0,1,1,0.55}
\definecolor{bluscuro}{rgb}{0.15, 0.2, .85}
\definecolor{bluchiaro}{cmyk}{1,.3,0.,0.1}
\definecolor{ForestGreen}{rgb}{0.13, 0.55, 0.13}
\definecolor{TLGreen}{RGB}{50, 164, 49}
\definecolor{TLOrange}{RGB}{231,180,22}
\definecolor{TLRed}{RGB}{204,50,50}

\newcommand{\TLBullet}[1]{\raisebox{-5pt}{\scalebox{0.23}{\begin{tikzpicture}\shadedraw[rounded corners=15pt, top color=gray!84!black,bottom color=black, line width=.6pt] (0,0) rectangle ++(6,2); \ifthenelse{#1=1}{\draw[fill=green,line width=1.pt]  (1,1) circle(.75cm);}{\draw[fill=green!35!black,line width=1.pt]  (1,1) circle(.75cm);}\ifthenelse{#1=2}{\draw[fill=yellow,line width=1.pt]  (3,1) circle(.75cm);}{\draw[fill=yellow!60!black,line width=1.pt]  (3,1) circle(.75cm);}\ifthenelse{#1=3}{\draw[fill=red,line width=1.pt]  (5,1) circle(.75cm);}{\draw[fill=red!50!black,line width=1.pt]  (5,1) circle(.75cm);}\end{tikzpicture}}}}
\newcommand{\apjl}{"Astrophys. J. Lett."}

\def\d{{\mathrm{d}}}

\newcommand{\bs}{\begin{subequations}}
\newcommand{\es}{\end{subequations}}

\newcommand{\be}{\begin{equation}}
\newcommand{\ee}{\end{equation}}
\renewcommand{\d}{{\rm d}}

\def\lsim{\mathrel{\rlap{\lower4pt\hbox{\hskip0.5pt$\sim$}}
    \raise1pt\hbox{$<$}}}         
\def\gsim{\mathrel{\rlap{\lower4pt\hbox{\hskip0.5pt$\sim$}}
    \raise1pt\hbox{$>$}}}         

\usepackage{siunitx}
\DeclareSIUnit \parsec {pc}
\DeclareSIUnit \arcsecondfull {arcsec}
\DeclareSIUnit \year{yr}
\DeclareSIUnit \day{day}
\DeclareSIUnit \hour{hr}
\DeclareSIUnit \radiant{rad}
\DeclareSIUnit \degfull{deg}
\DeclareSIUnit \erg {erg}
\DeclareSIUnit \Lsun {L_\odot}
\DeclareSIUnit \Msun {M_\odot}
\DeclareSIUnit \AstroUnit {au}

\usepackage{nicefrac}

\newcommand{\sapienza}{Dipartimento di Fisica, Sapienza Università di Roma \& INFN, Sezione di Roma, Piazzale Aldo Moro 5, 00185, Roma, Italy}

\newcommand{\sissa}{SISSA, Via Bonomea 265, 34136 Trieste, Italy \&  INFN Sezione di Trieste}

\newcommand{\ifpu}{IFPU - Institute for Fundamental Physics of the Universe, Via Beirut 2, 34014 Trieste, Italy}

\newcommand{\cern}{
CERN, Theoretical Physics Department,
Esplanade des Particules 1, Geneva 1211, Switzerland}

\begin{document}

\title{Cosmology and nuclear-physics implications of a subsolar gravitational-wave event}

\author{Francesco Crescimbeni}
\affiliation{\sapienza}

\author{Gabriele Franciolini}
\affiliation{\cern} 

\author{Paolo Pani}
\affiliation{\sapienza}

\author{Massimo Vaglio}
\affiliation{\sissa}
\affiliation{\sapienza}
\affiliation{\ifpu}


\begin{abstract}
\noindent
Detecting a compact subsolar object would have profound implications in physics, the reach of which depends on the nature of the object. Here, we explore such consequences for a putative subsolar-mass  gravitational wave event detected by the LIGO-Virgo-KAGRA Collaboration. 
We forecast that the nature of a subsolar binary (made of light neutron stars, primordial black holes, or more exotic compact objects) can be 
inferred with a great statistical confidence level already during the ongoing fourth observing run, based on the large tidal deformability effects on the signal.
The detection of a primordial black hole would have implications for cosmology and dark matter scenarios, while the measurement of the tidal deformability of a subsolar neutron star could rule out or confirm the existence of strange stars made of quarks.
\end{abstract}

\preprint{CERN-TH-2024-121}

\maketitle

\section{Introduction}
%
Since the groundbreaking detection of a gravitational wave (GW) event by LIGO in 2015~\cite{LIGOScientific:2016aoc}, the LIGO-Virgo-KAGRA~(LVK) Collaboration has reported nearly a hundred GW detections originating from the coalescence of compact binary systems~\cite{LIGOScientific:2018mvr, LIGOScientific:2020ibl, KAGRA:2021vkt}. Current GW interferometers are capable of identifying compact binary coalescences in a wide range of masses and are sensitive to mergers of hypothetical subsolar mass (SSM) compact objects (with mass $m < M_\odot$). Conventional formation models~\cite{Burrows:2020qrp} suggest that astrophysical compact objects, such as black holes~(BHs) and neutron stars~(NSs), should have masses exceeding the solar mass (see, e.g., Refs.~\cite{Sukhbold:2015wba,Ertl:2019zks,Burrows:2019rtd,Woosley:2020mze,Muller:2024aod} for recent simulations), although white dwarfs and NSs can theoretically be subsolar~\cite{Kilic:2006as, 1983bhwd.book.....S,Suwa:2018uni,Metzger:2024ujc}.

The detection of SSM compact objects could either indicate a novel formation mechanism beyond the standard-model stellar core-collapse scenarios~\cite{Metzger:2024ujc,Muller:2024aod} or imply evidence of new physics such as primordial black holes (PBHs)~\cite{Zeldovich:1967lct,Hawking:1971ei,Carr:1974nx, Carr:1975qj} (see, e.g., Refs.~\cite{Carr:2020gox,Green:2020jor} for recent reviews) or more exotic objects~\cite{Cardoso:2019rvt} such as Q-balls/boson stars~(BSs)~\cite{Coleman:1985ki,Colpi:1986ye, Liebling:2012fv}, fermion-soliton stars~\cite{Lee:1986tr,DelGrosso:2023trq,DelGrosso:2023dmv}, etc.
Whatever the case, such a discovery would have important implications for astrophysics, cosmology, and fundamental physics. 

Examples of SSM compact objects may have already been observed. HESS~J1731–347~\cite{2022NatAs...6.1444D} is a candidate NS with mass $0.77^{+0.20}_{-0.17} M_\odot$, while the candidate object reported in Ref.~\cite{2019ApJ...881L...3M} was possibly identified as a white dwarf with mass $0.20^{+0.01}_{-0.01} M_\odot$.

Observing SSM objects in compact binary coalescing systems is, therefore a primary target for current and future~\cite{Branchesi:2023mws} GW interferometers. 
Several GW searches for compact coalescing binaries with at least one SSM component have been performed using LVK data, but no conclusive detections have been made~\cite{Abbott_2018, Abbott_2019, Nitz_2021, Abbott_2022}. However, in the last concluded LVK observing run, O3b, three candidates of SSM binary BH events were identified~\cite{Abbott_2022}, and one candidate from the O2 data was reexamined in Ref.~\cite{Morras:2023jvb}. None of these candidates was confirmed due to their relatively high false alarm rates.

\begin{figure}[t!] 
\centering
\includegraphics[width=0.49\textwidth]{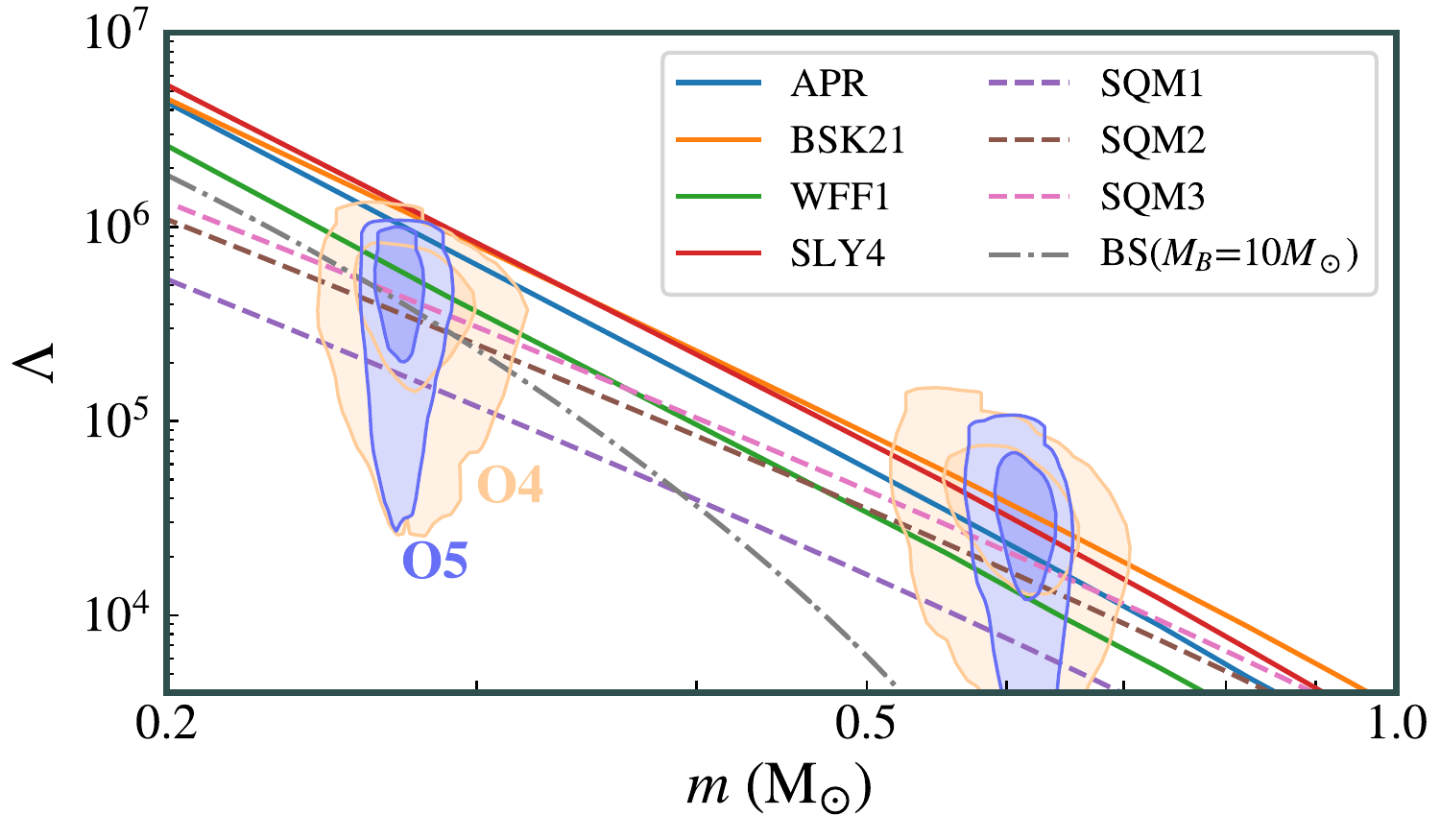} 
\caption{The posterior distribution ($50\%$ and $90\%$ C.L.) of mass and tidal deformability in a simulated observation of a SSM binary NS during O4 (O5) with ${\rm SNR}=25$ (${\rm SNR}=44$), assuming strange quark matter EoS SQM3. Different curves correspond to theoretical predictions for different NS EoS and for BSs.
As discussed below, such a measurement and Bayesian EoS selection~\cite{Pacilio:2021jmq} during O4/O5 would exclude the ordinary EoS with normal $npe\mu$ matter and would also exclude the PBH and BS origin of the binary.
}
\label{fig:LambdaM}  
\end{figure}

The scope of this work is to explore the following question: What would be the implications of a putative SSM GW event? As we will show, the ongoing LVK O4 run (and even more so the next O5 run scheduled to start in 2026) has sufficient sensitivity to discern among various scenarios, all with very impactful consequences.

In particular, we present the results of thorough Bayesian inferences on the candidate event SSM200308 and on projected O4 and O5 data, using several models for SSM objects, including PBHs, light NSs, and BSs, the latter taken as a representative model of exotic compact objects. We show that already SSM200308, if interpreted as a real event, would decisively exclude some models of NSs and BSs compared to the PBH hypothesis, based on the large tidal deformability effects on the waveform~\cite{Crescimbeni:2024asy,Golomb:2024mmt}.
We will show that in O4 this distinction would be crystal clear, confirming or ruling out a putative PBH event. In the latter case, using O4–O5 data it would be possible to use tidal-deformability measurements to decisively confirm/exclude light NSs with strange quark matter~\cite{Weber:2004kj,Haensel:1986qb}, giving a decisive contribution to the quest for the equation of state~(EoS) of nuclear matter at ultrahigh densities (see, e.g., Refs.~\cite{Prakash:1995uw, Lattimer:2000nx,Agathos:2015uaa}). As anticipated in Fig.~\ref{fig:LambdaM} and discussed below, this compelling opportunity is provided by the fact that, in the SSM range, quark stars have a much smaller radius (and hence much smaller tidal deformability) than ordinary-EoS NSs, and that overall the tidal deformability for SSM NSs is very large~\cite{Crescimbeni:2024asy} and so can be more easily  measured than for heavier NSs.

\section{Methodology and Models}
We use the same methodology implemented in Ref.~\cite{Crescimbeni:2024asy}, summarized below and detailed in Appendix~\ref{app:PE}.
In particular, the Bayesian inference on both real and synthetic GW data is done with the public software {\tt BILBY}~\cite{BILBY}. 
We consider quasicircular binaries, for which the waveform parameters common in all models are
\begin{equation}\label{eq:paramBBH}
{\boldsymbol \theta}
= \{m_1, m_2, d_L, \theta, \phi, \theta_{JN},  \psi, t_c, \Phi_c, \chi_{1}, \chi_{2} \}\,, 
\end{equation}
where $m_{1,2}$ are the source-frame masses, $\chi_{1,2}$ are the spin magnitudes, $\theta=\pi/2-\delta$ and $\phi$ the sky position coordinates (with $\phi$ and $\delta$ being the right ascension and declination, respectively), $d_L$ is the luminosity distance, $\theta_{JN}$ is the inclination angle of the binary with respect to the line of sight, $\psi$ the polarization angle, $t_c$ and $\Phi_c$ the time and phase of coalescence, respectively.

{
\renewcommand{\arraystretch}{1.2}
\setlength{\tabcolsep}{12pt}
\begin{table*}[t!]
\begin{tabular}
{ |>{\centering\arraybackslash}m{1.45cm}|>{\centering\arraybackslash}m{1.45cm}|>{\centering\arraybackslash}m{1.45cm}|>{\centering\arraybackslash}m{1.45cm}|>{\centering\arraybackslash}m{1.45cm}|>{\centering\arraybackslash}m{1.45cm}|>{\centering\arraybackslash}m{1.45cm}|}
\hline
\hline
Model  & BH1  & BH2 & Agnostic & NS1 & NS2 & BS
\\
\hline
\hline
$m_1(M_\odot)$ & $0.65^{+0.17}_{-0.15}$ & $0.72^{+0.20}_{-0.17}$ & $0.57^{+0.13}_{-0.10}$ & $0.59^{+0.29}_{-0.08}$ & $0.82^{+0.20}_{-0.14}$ & $0.50^{+0.10}_{-0.07}$\\
$m_2(M_\odot)$ & $0.26^{+0.07}_{-0.04}$ & $0.23^{+0.06}_{-0.04}$ & $0.29^{+0.05}_{-0.05}$ & $0.27^{+0.03}_{-0.08}$ & $0.21^{+0.03}_{-0.03}$ & $0.32^{+0.05}_{-0.05}$\\
$\chi_\text{\rm eff}$ & $0.41^{+0.05}_{-0.04}$ & $0.41^{+0.22}_{-0.05}$ & $-0.13^{+0.08}_{-0.09}$ & $0.15^{+0.16}_{-0.43}$ & $0.72^{+0.07}_{-0.26}$ & $0.36^{+0.25}_{-0.21}$\\
$\chi_\text{p}$ & $0.45^{+0.26}_{-0.26}$ & -  &  -  & - & - & - \\
$d_L [{\rm Mpc}]$ & $80^{+37}_{-29}$ & $83^{+41}_{-33}$ & $97^{+45}_{-41}$ & $110^{+139}_{-50}$ & $76^{+37}_{-28}$ & $106^{+84}_{-45}$ \\
$ {\Lambda_1}/{10^5} $ & - & - & $-4^{+15}_{-10}$ & $5^{+28}_{-3}$ & $6^{+8}_{-5}$ & - \\
$ {\Lambda_2}/{10^7} $ & - & - & $3^{+23}_{-12}$ & $1.3^{+0.6}_{-0.8}$ & $0.3^{+0.3}_{-0.3}$ & - \\
$ {\kappa_1}/{10^3} $ & - & - & $15^{+347}_{-351}$ & -  & - & - \\
$ {\kappa_2}/{10^3} $ & - & - & $-287^{+114}_{-120}$ & - & - & - \\
$\log_{10}\tilde \lambda_f$ & - & - & $-1.01^{+0.65}_{-0.42}$ & - & - & - \\
$M_B[M_\odot]$ & - & - & - & - & - & $10^{+2}_{-2}$ \\
\hline
\hline
$\log_{10}\mathcal{B}$& - & 0.31 & -1.64 & -2.68 & 0.22 & -2.26 \\
\hline
\hline
\end{tabular}
\caption{90\% C.I. inferred for the trigger event SSM200308 using different waveform models: BH1 (no tidal effects, precession), BH2 (no tidal effects, no precession), Agnostic (generic tidal deformability, quadrupoles, and tidal cutoff frequency), NS1/2 (two waveform models for light binary NS), and BS (a model for BS binaries). The Bayes factors are defined as $\mathcal{B}_i={\cal Z}_i/{\cal Z}_{{\rm BH1}}$, where ${\cal Z}_i$ is the evidence of a given model and $i=\{{\rm BH2, Agnostic, NS1, NS2, BS}\}$.
According to Jeffreys's scale criterion~\cite{Jeffreys}, a $\log_{10}$ Bayes factors greater than $1$ (respectively, $2$) would imply a strong (respectively, decisive) Bayesian evidence in favor of a given model relative to BH1.
}
\label{tab:results_open_data}
\end{table*}
}

We consider six different models:

i)~\textit{BH1}. The same IMRPhenomPv2 waveform~\cite{Hannam:2013oca} used in Ref.~\cite{Prunier:2023cyv} to analyze SSM200308. This model assumes a BH binary and includes spin precession, so it augments the parameters listed above with the spin angles. We adopt the same priors of Ref.~\cite{Prunier:2023cyv}, with extended ranges in some cases; see Appendix~\ref{app:PE} for more details.

ii)~\textit{BH2}. This is a simplified BH binary model based on the TaylorF2 waveform~\cite{Damour:2000zb}, including up to 3.5 post-Newtonian~(PN) terms~\cite{PhysRevD.44.3819,Damour:2000gg,PhysRevD.71.084008,Buonanno:2009zt} in the point-particle phase and up to 4PN spin effects (see Refs.~\cite{Franciolini:2021xbq,Favata:2021vhw} and references therein for details, e.g., Refs.~\cite{Kidder:1992fr,PhysRevD.48.1860,Kidder:1995zr,Poisson:1997ha,Mikoczi:2005dn,Gergely:1999pd,PhysRevD.74.104034,Mishra:2016whh}) assuming spins aligned with the orbital angular momentum.

iii)~\textit{Agnostic}. TaylorF2 waveform augmented with the 5PN and 6PN tidal terms in the phase~\cite{Wade:2014vqa}, parametrized in terms of the dimensionless tidal deformabilities $\Lambda_i$ ($i=1,2$) of the binary components~\cite{Crescimbeni:2024asy}, and with arbitrary spin-induced quadrupole terms at 2PN~\cite{Poisson:1997ha}, parametrized in terms of normalized quadrupole moments $k_i$ (with $k_i=1$ in the BH case). 
As discussed in Ref.~\cite{Crescimbeni:2024asy}, SSM binaries have high merger frequencies outside the bandwidth of current detectors. The observed signal originates from the early inspiral phase, where the TaylorF2 approximant remains accurate. This approach also avoids the need for more sophisticated binary NS waveform models, which are inaccurate in the SSM range~\cite{Markin:2023fxx}.
Finally, our model includes the tapering function introduced in Ref.~\cite{Crescimbeni:2024asy} (see also Ref.~\cite{DeLuca:2022xlz}) to account for tidal disruption of objects composing the binary, which is parametrized in terms of a cutoff frequency $f_\text{\tiny cutoff}=\tilde\lambda_f f_\text{\tiny ISCO}$, where 
$f_\text{\tiny ISCO}= 2.2\,  {\rm kHz} [M_\odot / (m_1+m_2)]$ is the frequency of the innermost stable circular orbit of the binary, and a slope $f_\text{\tiny slope}=\delta\tilde\lambda_f f_\text{\tiny ISCO}$, which controls the smoothness of the tapering, fixed to $\delta\tilde\lambda_f=10^{-2}$. In practice, this model augments BH2 with five extra parameters: \{$\Lambda_i$, $k_i$, $\tilde\lambda_f$\}.
As usually done, we perform the inference in terms of the effective parameters 
$\tilde\Lambda(\Lambda_i,m_i)$ and $\delta\tilde\Lambda(\Lambda_i,m_i)$, associated to the  5PN and 6PN order contribution to the GW
phase (see, e.g., Ref.~\cite{Crescimbeni:2024asy} for their explicit expressions). 
With the model being agnostic, we implement large uniform priors on the extra parameters (in particular $\Lambda_i$ and $k_i$ can be either positive or negative) and the same priors used in BH2 for the standard parameters.

iv)~\textit{NS1}. This model is based on Agnostic but enforces some relations among parameters, motivated by concrete NS models.
In particular, we implement the approximately EoS-independent ``Love-Q'' relations~\cite{Yagi:2013bca,Yagi:2013awa,Yagi:2016bkt,Silva:2016myw} which relate $k_i$ to $\Lambda_i$.
Finally, we assume a cutoff frequency given by the Roche overflow of the secondary star~\cite{Shibata:2001ag, Marronetti:2003hx,Dietrich:2015pxa, Bernuzzi:2020txg}.
Fitting the numerical simulations presented in Ref.~\cite{Bandopadhyay:2022tbi}, we estimate~\cite{Crescimbeni:2024asy}
\begin{align}\label{eq:RO}
    f_\text{\tiny cutoff} / {\rm Hz}
    =
    -26.9 - 35.5\left ( \frac{m_1}{M_\odot} \right)
    - 3.02\left ( \frac{m_1}{M_\odot}\right)^2 
    \nonumber 
    \\
    + 1690 \left (\frac{m_2}{M_\odot} \right ) 
    - 575 \left ( \frac{m_2}{M_\odot} \right )^2.
\end{align}
as a function of the binary masses. This fit reproduces the numerical results with errors $<10\%$ in the range of masses $[0.2,1] M_\odot$, and is not very sensitive to the EoS~\cite{Bandopadhyay:2022tbi}. 
In summary, this model has the same parameters of~BH2 plus the tidal deformabilities $\Lambda_1$ and $\Lambda_2$ that depend on the EoS and are kept as free parameters. 

v)~\textit{NS2}. Similar to NS1, but we ignore the Love-Q relations and assume the same spin-induced quadrupole moments of a BH ($k_i=1$).

vi)~\textit{BS}. This model implements a system of binary BSs with strong quartic interactions in the potential, for which the template was already developed in Ref.~\cite{Vaglio:2023lrd}. In this case, the parameters ${\boldsymbol \theta}$ are supplemented by $M_B$, which is the mass coupling constant related to the quartic potential~\cite{Pacilio:2020jza,Vaglio:2023lrd}. We also insert the waveform smoothing, which is controlled by the Roche frequency~\cite{Vaglio:2023lrd},
\begin{equation}
f_\text{\tiny cutoff} = \frac{{c^3}}{{\pi G (m_1+m_2)} }\sqrt{\frac{{(m_{1} + m_{2})^3}}{{m_{1} m_{2}^2}}} 
\left[
\frac{{\mathcal{C}(m_{2}/{M_B})}}{{g}}
\right]^{\frac{3}{2}},
\end{equation}
where $g\approx 2.44$ and compactness of this BS model is
\begin{equation}
C(m/{M_B}) \approx \frac{1}{{7.5 + 48.8 \left(1 - \frac{m}{0.06 M_B}\right)^2}}.
\end{equation} 
The tidal deformability is computed as in Refs.~\cite{Pacilio:2020jza,Sennett:2017etc,Vaglio:2023lrd}.

\section{Analysis of SSM200308}
With a false-alarm rate of 1 per 5 years, SSM200308 stands out as the most significant SSM candidate identified by the GstLAL pipeline~\cite{Cannon:2020qnf} in both LIGO Hanford and LIGO Livingston detectors. Although SSM200308 did not generate a trigger in Virgo with a signal-to-noise ratio (SNR) above the single detector threshold, Virgo was operational at that time, and its data are included in our analysis.

SSM200308 was initially analyzed under the assumption that it resulted from the merger of binary BHs, with reported masses of $m_1 = 0.62^{+0.46}_{-0.20}\,M_\odot$ and $m_2 = 0.27^{+0.12}_{-0.10}\,M_\odot$ at a redshift of $z = 0.02^{+0.01}_{-0.01}$ (90\% C.I.)~\cite{Prunier:2023cyv}. Recent forecasts, however, showed that the ongoing observing run O4 could potentially distinguish, through tidal deformability measurements, whether the trigger originated from the merger of two PBHs or other SSM candidates~\cite{Crescimbeni:2024asy}.
Motivated by these findings, we conduct Bayesian inferences on the SSM200308 data under the different hypotheses presented above. 
Our results do not alter the low significance of SSM200308, and we do not claim evidence of this being an actual event, as it may also have originated from environmental or instrumental noise. Rather, we show that such long-duration and low-mass signals could give valuable insight to distinguish different hypothesis, even more so
with the anticipated improvements in sensitivity of future observing runs.

The results of our analysis are summarized in Table~\ref{tab:results_open_data}, where we show the $90\%$ C.I. for the most representative parameters. 
The posterior distributions of the masses are compatible in all models and at least the secondary mass $m_2$ could always be statistically inferred to be subsolar and to satisfy the condition $m_2 + 3 \Delta m_2 < M_\odot$~\cite{Franciolini:2021xbq}, where $\Delta m_2$ is its standard deviation.
The results with BH1 are compatible with the ones reported in Ref.~\cite{Prunier:2023cyv}. We adopt this as the baseline to compare the Bayes factors between different models. 
For BH1,
we find that the data prefer a positive effective spin $\chi_\text{\rm eff}$ with high significance, while the precession parameter $\chi_{\rm p}$ is only weakly constrained.
Assuming the BH2 model, which neglects spin precession, we find similar results to the  BH1 case. 
Therefore, given also the small Bayes factor in the table ($\log_{10}{\cal B}=0.31$), we conclude there is limited evidence for precession in current data.

With the Agnostic hypothesis, we observe that the posteriors of the dimensionless quadrupoles $k_i$ are broad, which is compensated by a low $\chi_{\rm eff}$, so that the corresponding 2PN phase term, controlled by the combinations $ k_i \chi_i^2$, remains small. 
Also, the tidal deformabilities are large enough to include all possible origins of the SSM candidate. 
This can be due to the limited sensitivity at large frequencies during the O3 run as well as the strong preference for a cutoff frequency below the ISCO, as indicated by the tapering parameter $\log_{10}\tilde \lambda_f$ being inferred to be smaller than zero with large significance. While the latter would disfavor the binary BH interpretation, the relative $\log_{10}$ Bayes factor to the BH1 case is negative, possibly due to weakly informative data and the larger number of parameters in the Agnostic model. 
For what concerns the models NS1 and NS2, while the latter has a Bayes factor similar to BH1, the former shows a decisive negative difference according to Jeffreys's scale~\cite{Jeffreys}. We expect that the inclusion of the Love-Q relations is responsible for generating instabilities at high spins. Note that, despite NS2 not being statistically disfavored, it infers highly spinning binary components which could be in tension with a NS interpretation, since old NSs in the late stages of the inspiral are expected to rotate slowly~\cite{Tauris:2017omb}.

Finally, under the BS hypothesis, one would place the coupling  $M_B$ close to $10M_{\odot}$, but also this scenario is decisively disfavored by the data compared to other models, probably for the same reason of NS1 as in the BS model we implement its own Love-Q relations~\cite{Vaglio:2022flq,Vaglio:2023lrd}.

\section{Forecasts for LVK O4/O5}
We now show the prospects of pinpointing the nature of a SSM object in the ongoing (O4) and next (O5) LVK runs.
We focus on the BH2 and NS2 models, which have similar evidence in the SSM2003308 analysis (see Table~\ref{tab:results_open_data}). We perform four injection$\rightarrow$recovery simulations: 
i)~BH2$\rightarrow$BH2,
ii)~BH2 $\rightarrow$NS2,
iii)~NS2$\rightarrow$ BH2, and
iv)~NS2$\rightarrow$NS2.
We assumed NS2 to be more conservative but checked that NS1 would give very similar results. In addition, including the Love-Q allows for better measurement of the binary spins.
For each run, we sample the parameters in Eq.~\eqref{eq:paramBBH}, with the addition of $\Lambda_i$ when performing an NS2 recovery, and assume a frequency range $f\in[10,2048]\,{\rm Hz}$, as done in Ref.~\cite{Crescimbeni:2024asy}. 
As a reference, we simulate a binary with the maximum likelihood parameters of SSM200308~\cite{Prunier:2023cyv}, 
at a distance either corresponding to ${\rm SNR}=25$ in O4 (i.e. ${\rm SNR}\approx 15$ in O3), or ${\rm SNR}=15$ in O4.
These two choices conservatively correspond to a luminosity distance $d_L=51\, {\rm Mpc}$ or $d_L=89\, {\rm Mpc}$, respectively.

\begin{figure} [!t] 
\centering  
\includegraphics[width = 1.02\linewidth]{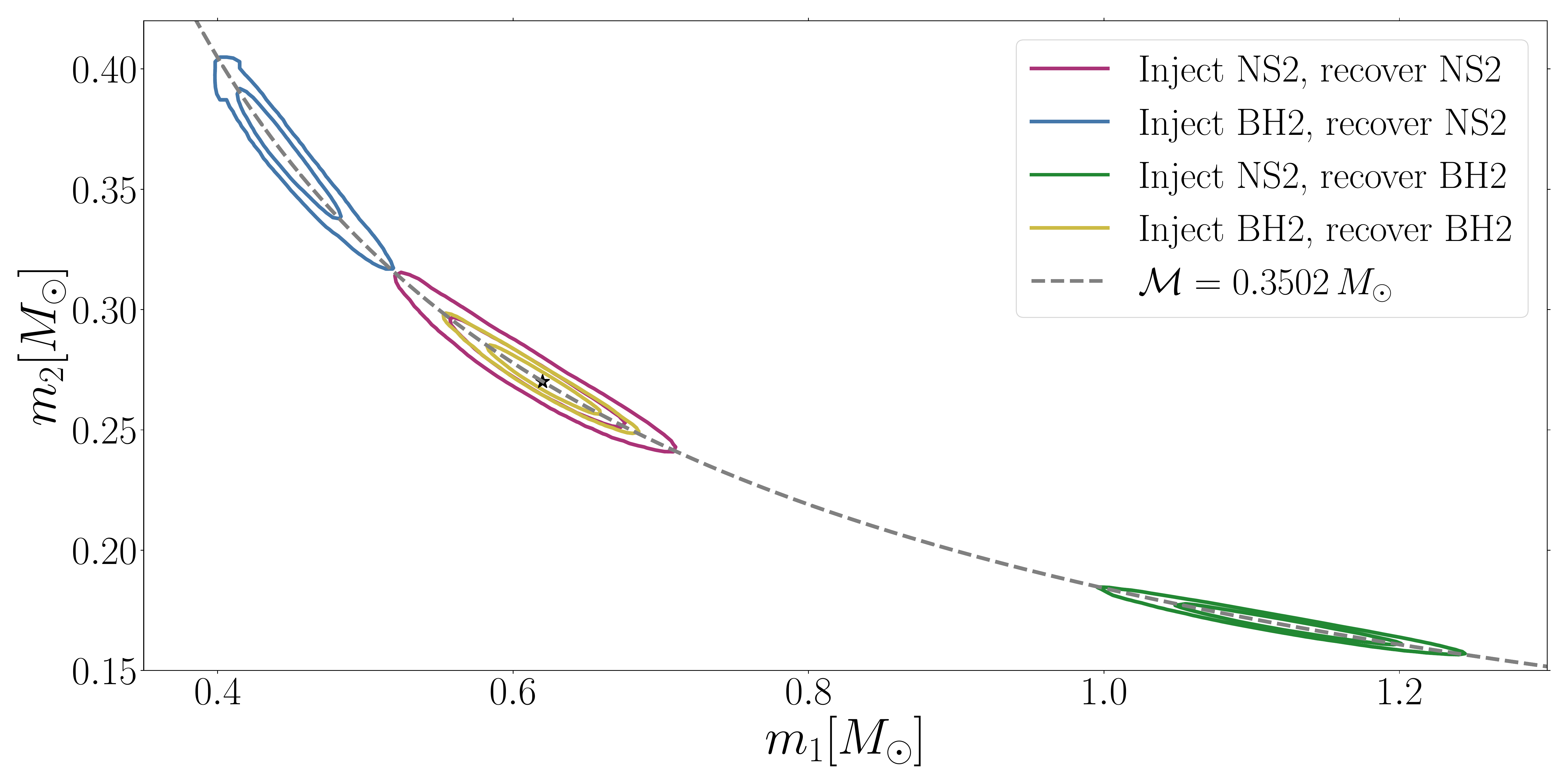} 
\includegraphics[width = 1.02 \linewidth]{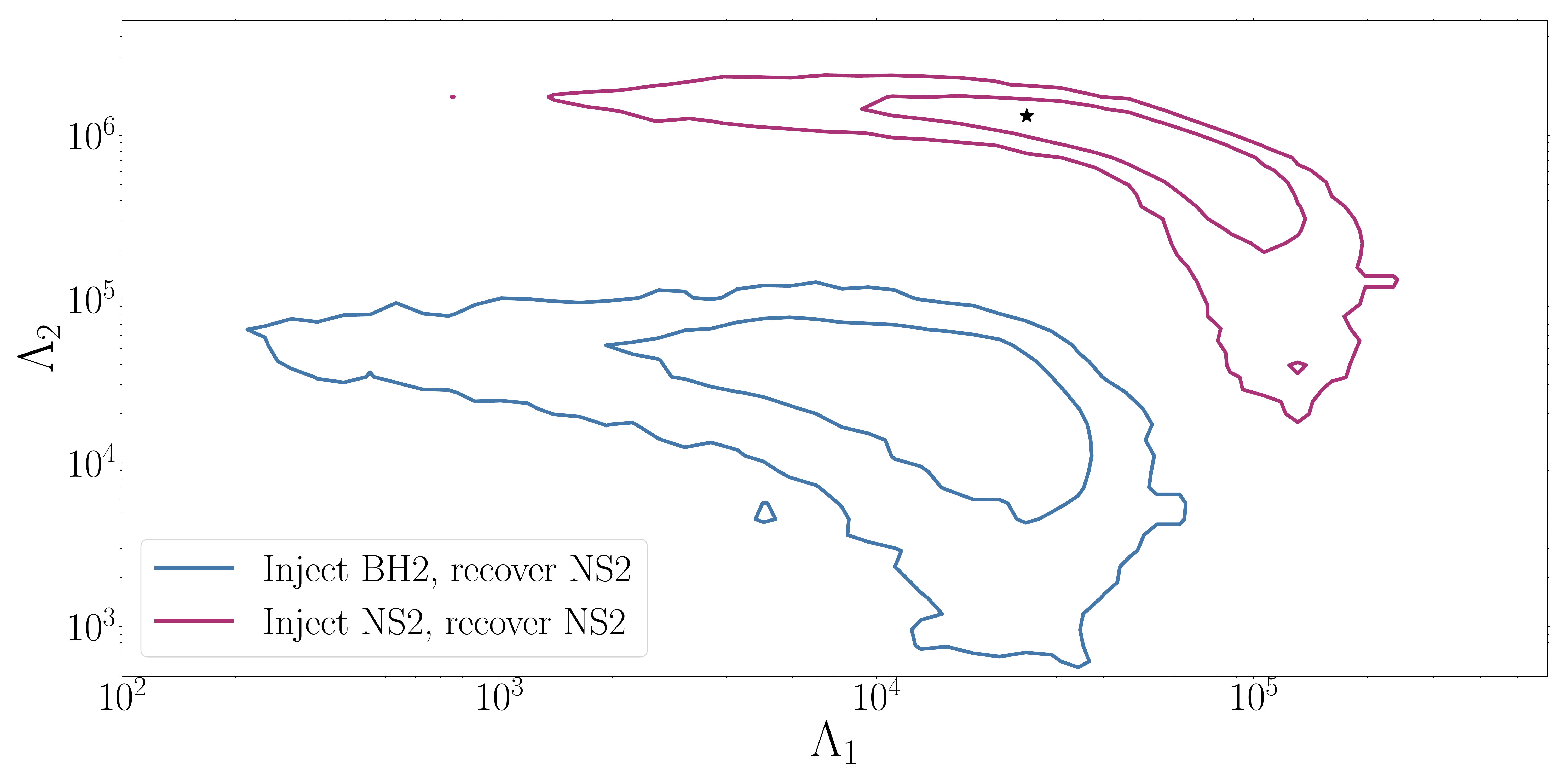}  
\caption{{\it Upper panel:} posterior distribution of the masses for the O4 simulations. The dashed line corresponds to fixing the chirp mass ${\cal M}$ to the injected value, showing that fitting with the incorrect model mostly biases the inferred symmetric mass ratio. 
{\it Lower panel:} same but for the tidal parameters. The black marker locates the injected values.
}
\label{fig:biasO4}  
\end{figure}

In Table~\ref{tab:results_O4}, we show the corresponding Bayes factors, focusing on two models (BH2 and NS2), which could not be discriminated according to the Bayesian analysis shown in Table~\ref{tab:results_open_data}.
We observe that O4 sensitivity is sufficient to distinguish a PBH signal from the NS one with similar masses, especially at higher SNR.
Our results strongly suggest that, should a PBH SSM binary be detected already in O4, it can be confidently distinguished from an SSM NS binary and vice versa. Because the two cases are already distinguishable in O4, O5 will be even more capable of discriminating between them. As an estimate, a SNR of 15 in O4 corresponds roughly to an SNR of 25 in O5 \cite{Crescimbeni:2024asy}.
Similar conclusions apply to the BS model (not shown), which can be even more strongly excluded in the case of a BH or NS injection.
Furthermore, as shown in Fig.~\ref{fig:biasO4}, interpreting a SSM BH binary signal with the wrong NS model (or vice versa) would introduce a significant bias in both masses and tidal deformabilities. 
This shows that performing model selection for these sources is crucial to properly identifying the nature of the objects.

\begin{table}[h]
\centering
\begin{tabular}{|>{\centering\arraybackslash}m{2.5cm}|>{\centering\arraybackslash}m{2.5cm}|>{\centering\arraybackslash}m{2.5cm}|}
\hline
\hline
Detector & $BH2$ $\rightarrow $ $NS2$  & $NS2$ $\rightarrow $ $BH2$ \\
\hline \hline
SNR=15 & -3.1  & -2.9 \\
SNR=25  & -5.1 & -11.0 \\
\hline
\hline
\end{tabular}
\caption{
$\log_{10}$ Bayes factors to assess the PBH or NS origin of a putative SSM event, normalized to the case where the injection and recovery assume the same model (i.e., negative values disfavor the wrong hypothesis). 
We assumed O4 sensitivity, and results for O5 at the same SNR would be very similar.
}
\label{tab:results_O4}
\end{table}

\section{Cosmology and Nuclear Physics implications}
Having assessed that the nature of a SSM can be confidently identified, we now turn our attention to the implications of such putative detection.

\subsection{Primordial hypothesis}
In the case of a SSM BH binary detected during O4, our results show that one can rule out the NS interpretation with decisive Bayesian evidence, mostly through the observation of the GW signal beyond the expected disruption frequency (i.e. $\tilde \lambda_f \sim 1$).\footnote{We are neglecting here the possible impact of environmental effects which could generate nonzero tidal effects in the relatively low-frequency range of the inspiral signal \cite{Baumann:2018vus, DeLuca:2021ite, DeLuca:2022xlz,Cardoso:2019upw, Cardoso:2021wlq,Cannizzaro:2024fpz,DeLuca:2024uju}.
These would yield frequency-dependent features that can be disentangled from those of NS tidal deformabilities.
} 
One could then infer the corresponding PBH abundance $f_{\rm PBH}$ (defined as the fraction of dark matter energy density made of PBHs) that controls the merger rate of similar SSM events. Following the analysis of Ref.~\cite{Franciolini:2022tfm}, based on a model saturating the GWTC-3 upper bound, one would need an abundance of order $f_{\rm PBH} \sim {\cal O}(10^{-2})$ to explain such a SSM event, a value that depends on the possible assumed mass distribution.
This is close but not excluded by the most stringent bounds in this mass range arising from missing microlensing signatures in OGLE datasets \cite{Mroz:2024mse}. Because of the finite width of the PBH mass distribution required at least by the features of the critical collapse, one could then expect PBH events also to contaminate the stellar mass range.
While such observation would point toward an important role of PBHs in the evolution of large-scale structures (see, e.g., Refs.~\cite{Carr:2018rid,Boldrini:2019isx,Inman:2019wvr,DeLuca:2020jug,Liu:2023pvq}), it would also strengthen bounds from indirect detection of particle dark matter, which is largely incompatible with an abundant population of compact objects boosting the annihilation J-factors~\cite{Lacki:2010zf,Eroshenko:2016yve,Boucenna:2017ghj,Adamek:2019gns,Bertone:2019vsk}.
Finally, the existence of a PBH population would require specific setups in the early Universe, either deviating from the vanilla slow-roll inflationary scenario~\cite{Motohashi:2017kbs} or other formation mechanisms based on new physics (see, e.g., Ref.~\cite{Flores:2024eyy}).
Another nonprimordial explanation could be
subsolar BHs formed from stellar transmutation triggered by dark matter accretion~\cite{Takhistov:2017bpt,Kouvaris:2018wnh,Takhistov:2020vxs,Dasgupta:2020mqg,Singh:2022wvw,Bhattacharya:2023stq,Chakraborty:2024eyx}, which anyway require new physics.

\subsection{NS equation of state}
If the SSM objects are identified as light NSs, they would point toward nonstandard formation scenarios, for example, supernovae occurring in sufficiently dense neutron-rich environments~\cite{Metzger:2024ujc},
since supernova theory struggles to explain subsolar NSs~\cite{Muller:2024aod}. 

Furthermore, the large tidal effects can be exploited to constrain the NS EoS. This was anticipated in Fig.~\ref{fig:LambdaM}, showing the projected constraints on $\Lambda_i$ and $m_i$ inferred from an injection using NS2 with a typical O4/O5 SNR. 
We selected several widely used representatives from different EoS classes commonly employed to model NS tidal deformability. Specifically, APR~\cite{Akmal:1998cf} and WFF1~\cite{Wiringa:1988tp} are based on microscopic models incorporating realistic nucleon-nucleon interactions, while SLy4 \cite{Douchin:2001sv} represents mean-field models with effective nuclear interactions, and BSk21 \cite{Goriely:2010bm} is a phenomenological model fitted to nuclear mass data. For strange quark matter (SQM) stars, we chose SQM3~\cite{Prakash:1995uw} as a representative model because its tidal deformability results are comparable to those of nonquark families. Notably, SQM1 and SQM2 share the same EoS as SQM3, differing only in the bag constant which only sets a different scale. The models also exhibit varying levels of stiffness: APR and BSk21 are relatively stiff, WFF1 and SLy4 are softer, and SQM3 represents a stiff quark matter EoS.

As a proof of principle, 
as shown in Fig.~\ref{fig:LambdaM}, we assume an EoS for SQM3 compatible with NSs with mass up to $\approx 2 M_\odot$ (as required by pulsar observations~\cite{Fonseca:2021wxt}). In the SSM range, this EoS predicts NSs with significantly smaller tidal deformability and radius than ordinary EoS produced by $npe\mu$ matter. Thus, if the event is generated by SSM strange stars,
the upper bounds on $\Lambda_2$ (i.e. the lighter object) are already tight enough in O4 (and especially O5) to disfavor ordinary EoS at more than $90\%$ confidence level, while the bounds on $\Lambda_1$ (i.e. the heavier object) are less constraining. 
Notice that the lower bounds Fig.~\ref{fig:LambdaM} (and Fig.~\ref{fig:biasO4}, bottom panel) are influenced by a volume effect, since uniform priors on $\Lambda_i$ tend to overestimate the posterior at larger values, and the simulated analyses are still prior dependent due to the relatively low SNR. To mitigate this, one could instead adopt log-uniform priors for $\Lambda_i$. Our choice remains conservative when estimating the upper bounds.

\begin{table}[]
\centering
\begin{tabular}{|>{\centering\arraybackslash}m{2.0cm}|>{\centering\arraybackslash}m{1.2cm}|>
{\centering\arraybackslash}m{1.2cm}|>{\centering\arraybackslash}m{1.2cm}|>
{\centering\arraybackslash}m{1.2cm}|>
{\centering\arraybackslash}m{1.2cm}|}
\hline
\hline
Detectors & $m_1[M_{\odot}]$ & $\texttt{APR}$ $\rightarrow $ $\texttt{SQM3}$  & $\texttt{SQM3}$ $\rightarrow $ $\texttt{APR}$ & $\texttt{WFF1}$ $\rightarrow $ $\texttt{SQM3}$  & $\texttt{SQM3}$ $\rightarrow $ $\texttt{WFF1}$ \\
\hline \hline
O4,~${\rm SNR}=25$ & $0.63$ & $-1.9$  & $-3.8$ & $0.1$ & $-0.4$ \\
 & $0.27$ & $-10.2$ & $-19.9$ & $-2.7$ & $-5.0$ \\
\hline
O5,~${\rm SNR}=44$ & $0.63$ & $-7.0$ & $-12.3$ & -0.2 & $-1.0$ \\
 & $0.27$ & $-37.5$ & $-88.8$ & $-11.3$ & $-25.1$ \\
\hline
\hline
\end{tabular}
\caption{
Same as Table~\ref{tab:results_O4}, but for different EoS in NS2. 
The source luminosity distance is $d_L=51\,{\rm Mpc}$ ($d_L=36\,{\rm Mpc}$) for $m_1=0.63M_\odot$ ($m_1=0.27M_\odot$).
For all the runs, we assume that $m_2=0.27M_{\odot}$. 
}
\label{tab:results_APR_vs_SQM3}
\end{table}

Distinguishing strange quark stars from ordinary NS supported by soft matter is more challenging, since in this case the predicted smaller $\Lambda_2$ can be compatible with the upper bounds.
This is the case of WFF1 and marginally of APR, which indeed cannot be excluded solely from the bounds shown in Fig.~\ref{fig:LambdaM}.
However, note that the analysis in Fig.~\ref{fig:LambdaM} is conservative since the inference is performed considering $\Lambda_1$ and $\Lambda_2$ as independent parameters, while in reality, for a given EoS, they are fixed in terms of the binary masses.
To obtain stronger constraints, one could perform a Bayesian model selection between two specific EoS models, implementing the EoS-dependent relations $\Lambda_i=\Lambda_i(m_i)$ in the waveform to reduce the number of parameters, as done in Ref.~\cite{Pacilio:2021jmq} for ordinary NSs.
The results of this analysis are presented in Table~\ref{tab:results_APR_vs_SQM3}, showing the Bayes factors for two injections (with different mass ratios) assuming APR or WFF1 EoS and recovery with SQM3 and vice versa.
Also, in this case, the results were obtained assuming NS2 but would be very similar for the less conservative model NS1.
A reference detection in O4 would already decisively exclude APR if the signal was generated by strange stars. However, the opposite case is less constraining, because the tidal deformability of APR is larger than that of SQM3, and the \emph{lower} bounds shown in Fig.~\ref{fig:LambdaM} are not stringent.
Remarkably, however, the same putative event detected during O5, or a comparable-mass light SSM binary in O4, would allow distinguishing between APR and SQM3 with exceptionally high statistical confidence.
Excluding WFF1 requires higher SNR or light comparable-mass SSM binaries, since in this case the tidal deformability is more similar to that of SQM3. One would need O5 sensitivity (corresponding to SNR = 44) and sufficiently light SSM components to exclude WFF1 from SQM3 with high confidence.

\begin{figure}[th!] 
\centering
\includegraphics[scale=0.58]{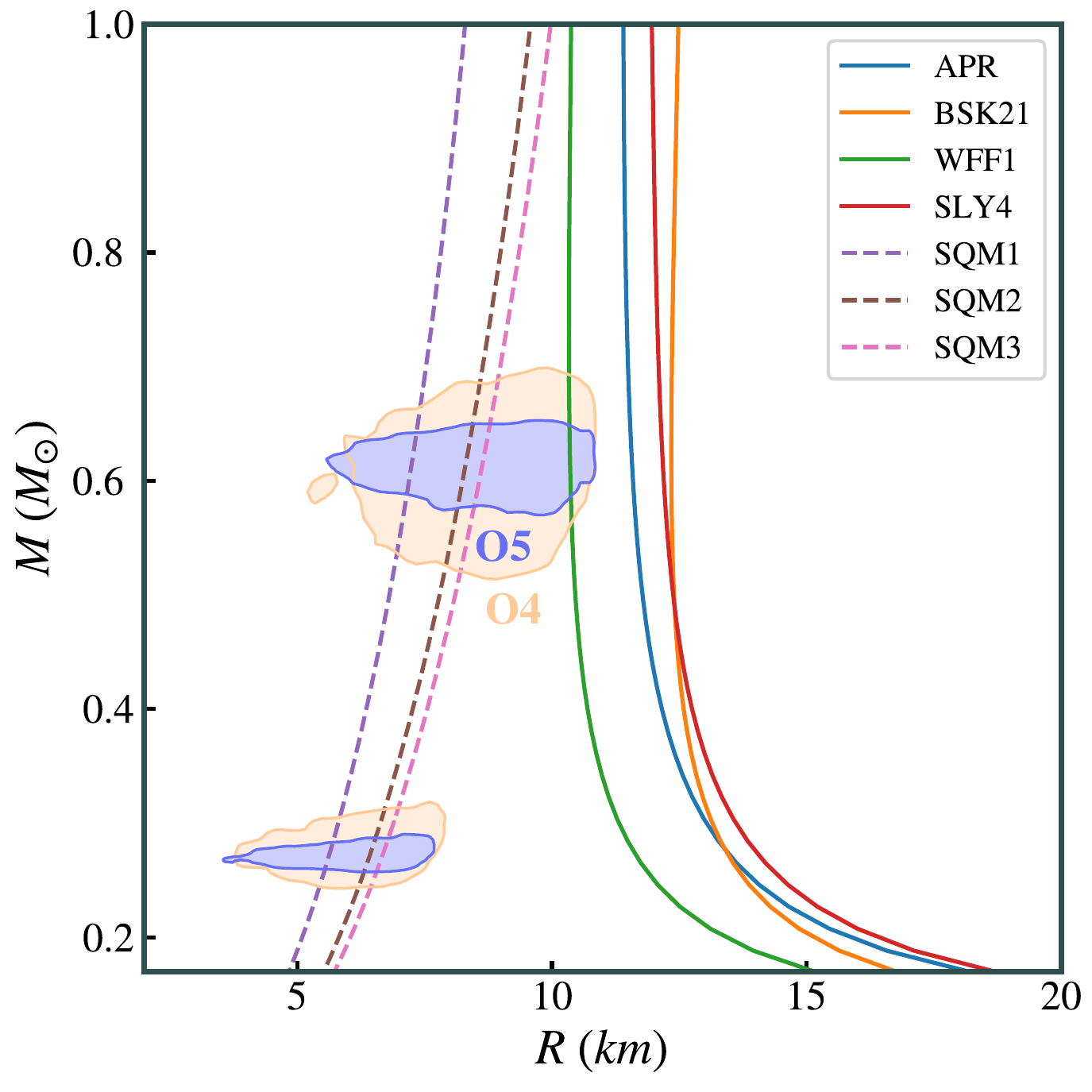} 
\caption{
The projected mass and radius of a subsolar mass NS obtained from the measurement of the tidal deformability.
We use $m_1= 0.63 M_\odot$ and $m_2 = 0.27 M_\odot$ with the SQM3 EoS detected in O4 and O5 with SNR = 25 and SNR = 44, respectively. 
Assuming a relationship for $\Lambda$ and the
star’s compactness valid for SQM1-3, we obtain the 90\% contours by mapping the posteriors of $\Lambda_i$ and $m_i$ into those of the radii of the binary components.
}
\label{fig:MR}  
\end{figure}

Overall, we conclude that a putative SSM event detected during the LVK O5 run would allow ruling out or confirming the existence of stars made of quarks, a key result toward the quest for the EoS of high-density nuclear matter~\cite{Prakash:1995uw, Lattimer:2000nx,Agathos:2015uaa}. This would translate into very stringent constraints on the NS masses and radii, as summarized in Fig.~\ref{fig:MR}, showing a scarcely constrained region of the typical mass-radius diagram of a NS (since there are currently no observational constraints on the radius of subsolar NSs).
Since strange stars have significantly smaller radii than ordinary NSs in the SSM range, the constraints on the mass-radius diagram are very informative to discriminate among different EoS.

\section{Conclusions}
We have explored the implications of a putative SSM event detectable by current GW interferometers.
Such detection would be a milestone for astrophysics and cosmology, potentially providing the first GW evidence for new physics. 
We showed that the past O3 LVK sensitivity was insufficient to confidently distinguish between different SSM models. However, the situation will drastically change during the ongoing O4 run and, especially, during the future O5 run. We showed that for a typical SSM event detected in O4, one can confidently identify the nature of the objects, being BHs, light NSs, or more exotic BSs. In case data favor a BH interpretation, this would likely imply that PBHs comprise a sizable fraction of dark matter and that also some stellar-mass mergers in LVK data should be primordial.
On the other hand, in case data favor the light NS hypothesis, O5 sensitivity would allow placing stringent upper bounds on the tidal deformability, which could be used to confirm or rule out NSs made of strange quark matter.
Overall, our results show that current interferometers have the potential to unveil exciting implications of a SSM detection, and we hope this potential will be fulfilled in the near future.

%

\section*{Acknowledgments}
We would like to thank V. De Luca, A. Iovino and A. Riotto for the fruitful discussions. Inference analyses have been performed at the Vera cluster supported by the Italian Ministry of Research and Sapienza University of Rome.
This work is partially supported by the MUR PRIN Grant No. 2020KR4KN2 ``String Theory as a bridge between Gauge Theories and Quantum Gravity'' and by the FARE programme (GW-NEXT, CUP:~B84I20000100001). M.V. acknowledges support from the PRIN 2022 Grant No. “GUVIRP
- Gravity tests in the UltraViolet and InfraRed with Pulsar timing”.

Inference simulations have been carried out with {\tt BILBY} \cite{Ashton:2018jfp}, a python-based software for gravitational wave inference. The manuscript content has been derived using publicly available software: {\tt matplotlib}, {\tt corner}, {\tt json}, {\tt numpy} \cite{Hunter:2007, corner, bray2014javascript, harris2020array}. Codes are available upon request and can be requested from \url{francesco.crescimbeni@uniroma1.it}.

\newpage
\appendix

\section{Parameter estimation}  \label{app:PE}

We provide here some additional details on the technical aspects of parameter estimation analyses performed in this work. 
We are interested in estimating the posterior distribution $p({\boldsymbol \theta}|s)$ of a set of hyperparameters ${\boldsymbol \theta}$, conditioned by the detection of a total signal
\begin{equation}
s(t)=h(t,{\boldsymbol \theta})+n(t)
\label{output_s}
\end{equation}
where $h(t,{\boldsymbol \theta})$ is the GW signal, and $n(t)$ is the stationary noise component due to the interferometer(s). 
The posterior distribution can be approximated as
\be\label{pos_dist_F}
p ({\boldsymbol \theta}| s) \propto \pi ({\boldsymbol \theta}) e^{-\frac{1}{2}(h ({\boldsymbol \theta}) - s|h ({\boldsymbol \theta}) - s)}
\ee
in terms of the prior distribution $\pi ({\boldsymbol \theta})$. The inner product adopted in Eq.~\eqref{pos_dist_F} is defined as
\be\label{innprod}
(g|h) = 2\int_{f_\text{\tiny min}}^{f_\text{\tiny max}} \d f \frac{\tilde h (f) \tilde g^* (f) + \tilde h^*(f) \tilde g(f)}{S_n(f)}\,,
\ee
in terms of the Fourier-transformed quantities and the detector noise power spectral density, $S_n(f)$. The frequency band $[f_\text{\tiny min},f_\text{\tiny max}]$ of interest depends on the specific detector. The SNR is given by ${\rm SNR}=\sqrt{(h|h)}$~\cite{bayes1, bayes2, bayes4}.

Bayes's theorem states then that
\begin{equation}
    p({\boldsymbol \theta}|s)=\frac{\mathcal{L}(s|{\boldsymbol \theta})\pi({\boldsymbol \theta})}{\mathcal{Z}(s)}\,,
    \label{post}
\end{equation}
where we defined:
\begin{itemize}[leftmargin=*]
    \item $\mathcal{L}(s|{\boldsymbol \theta})$ is the probability of having a signal $s$ given the (source) parameters ${\boldsymbol \theta}$ and is known as the likelihood function; the choice of the likelihood is linked to the noise model that we adopt, for instance, a Gaussian one (see, e.g., Ref.~\cite{Maggiore:2007ulw}).
    \item $\pi({\boldsymbol \theta})$ indicates the {prior probability distribution} of having the set of parameters ${\boldsymbol \theta}$; it represents our knowledge about ${\boldsymbol \theta}$ before we make the measurement.
    \item $\mathcal{Z}(s)$ is the {evidence}, or marginal likelihood, that is,
\begin{equation}
\mathcal{Z}(s)=\int\mathcal{L}(s|{\boldsymbol \theta})\pi({\boldsymbol \theta})
d {\boldsymbol \theta} 
\end{equation}
where the integral is intended over the full $n$-dimensional parameter space.
\end{itemize}
We carried out Bayesian inference with the public software {\tt BILBY}~\cite{BILBY}, which makes use of \textit{DYNESTY} nested sampling~\cite{dynesty}. To speed up the simulations, we implemented the relative binning technique~\cite{Zackay:2018qdy}, which greatly speeds up the likelihood evaluation by expanding it over some fiducial parameters. 
For the real-data part, the relative binning also requires a set of fiducial parameters~\cite{Krishna:2023bug}, which are the ones that are meant to maximize the likelihood. Since these are not known \textit{a priori}, we have obtained maximum likelihood estimates by iterating an optimization algorithm that makes use of routines in the \textit{SciPy} library. To compute the Bayes factors of different models, we adopted the same broad priors for all inferences (see the next section). To speed up the analysis, the only parameter for which we assume a narrow prior is the chirp mass, which is always well-measured.

\section{Parameter priors}

We report the priors adopted in our inference analyses in Table~\ref{tab:priors}.

\begin{table}[h]
\centering
\begin{tabular}{|>{\centering\arraybackslash}m{2.7cm}|>{\centering\arraybackslash}m{2.7cm}|>{\centering\arraybackslash}m{2.7cm}|}
\hline \hline
Parameter            & Lower boundary            & Upper boundary            \\ \hline \hline
$\mathcal{M}\, [M_\odot]$        & 0.351                  & 0.355                  \\ 
$\eta$               & 0.10                   & 0.25                   \\ 
$\tilde \Lambda$     & -$10^6$        & $10^6$         \\ 
$\delta\tilde \Lambda$ & -$1.5 \cdot 10^7$    & $1.5 \cdot 10^7$       \\ 
$\chi_1$             & -0.9                   & 0.9                    \\ 
$\chi_2$             & -0.9                   & 0.9                    \\ 
$\kappa_1$           & -$0.5 \cdot 10^3$      & $0.5 \cdot 10^3$       \\ 
$\kappa_2$           & -$0.5 \cdot 10^3$      & $0.5 \cdot 10^3$       \\ 
$d_L\,[{\rm Mpc}]$                & 5                      & 300                    \\ 
$\theta_\mathrm{JN}$ & 0                      & $\pi$                  \\ 
$\phi$               & -$\pi$                 & $\pi$                  \\ 
$\delta$             & -${\pi}/{2}$       & ${\pi}/{2}$        \\ 
$\alpha$             & 0                      & $2\pi$                 \\ 
$\psi$               & 0                      & $\pi$                  \\ 
$t_\mathrm{geocent}$ & $t_{\rm trigger} - 0.1 s$    & $t_{\rm trigger}+ 0.1  s$    \\
\hline \hline
\end{tabular}
\caption{Parameters adopted in the analysis.
All priors are defined to be uniform within the reported boundaries, except for $d_L$ (uniform in source frame), $\delta$ (cosine), and $\theta_{JN}$ (sine). We further used the trigger time $t_{\rm trigger}=1267725971.0234s$ reported in Ref.~\cite{Prunier:2023cyv} for SSM200308.}
\label{tab:priors}
\end{table}

\section{Fits for tidal deformabilities}

We compile here the fits utilized throughout this work. In deriving the Bayes factors presented in Table~III and in performing EoS selection with the NS2 model, we connected the tidal deformability to the masses through the approximated relation
\begin{equation}
\log_{10} (\Lambda) = a_0 + a_1 \log_{10} (m) + a_2 \log_{10}^2 (m)
\label{eq:lambda_fit}
\end{equation}
whose coefficients have been numerically fitted for each EoS and are reported in Table~\ref{tab:Eos_fit_coeffs}. The fits are accurate within $10\%$ in the mass range $\in [0.2,1]M_\odot$, with an average error below $2\%$. 

To map the posterior into the mass-radius plane, as shown in Fig.~3, we used the following expression to approximate the relationship between a strange star’s compactness and its tidal deformability:  
\begin{equation}
C(\Lambda) = a \cdot \Lambda^{-\frac{2}{5}} + b \cdot \Lambda^{-\frac{1}{5}} + c\,,
\end{equation}
where
\begin{align*}
a &= -7.05845 \times 10^{-1} \\
b &= 8.46607 \times 10^{-1} \\
c &= 7.60135 \times 10^{-4}
\end{align*}
Because the SQM1-3 models only differ by a dimensional scale factor~\cite{Prakash:1995uw}, the correspondence between dimensionless quantities such as $C$ and $\Lambda$ remains unaltered when changing from one SQM EoS to another so that the same coefficients describe the three cases. The accuracy of the fit in the range of masses $\in [0.2,1]M_\odot$ is within $0.5\%$.

\begin{table}[h]
\centering
\begin{tabular}{|>{\centering\arraybackslash}m{2cm}|>{\centering\arraybackslash}m{2cm}|>{\centering\arraybackslash}m{2cm}|>{\centering\arraybackslash}m{2cm}|}
\hline
\hline
EoS & $a_0$ & $a_1$ & $a_2$ \\
\hline \hline
APR & 3.25832 & -5.09578 &  -0.40728\\
WFF1 & 3.05497 & -4.97712 &  -0.30354 \\
SQM3 & 3.40331 & -4.25522 &  -0.49287 \\
\hline
\hline
\end{tabular}
\caption{Coefficients in Eq.~\eqref{eq:lambda_fit} for the EoS considered in Table~III.}
\label{tab:Eos_fit_coeffs}
\end{table}

\bibliography{main}
\let\addcontentsline\oldaddcontentsline

\end{document}
